\documentclass{article}

\usepackage{arxiv}

\usepackage{cite}
\usepackage{graphicx}
\usepackage{multirow}
\usepackage{amsmath}
\usepackage[version=3]{mhchem}
\usepackage{stfloats}
\usepackage{algorithm}  
\usepackage[noend]{algpseudocode}
\usepackage{url}
\usepackage[utf8]{inputenc}
\usepackage[T1]{fontenc}
\usepackage{mathptmx}
\usepackage{txfonts}
\usepackage{comment}
\usepackage{soul}
\usepackage[normalem]{ulem}
\usepackage{color}
\usepackage{xcolor}
\usepackage[bookmarksnumbered=true, hidelinks]{hyperref}
\usepackage{braket}
\usepackage{booktabs}
\usepackage{wrapfig}

\title{Experimental Demonstration of Software-Orchestrated Quantum Network Applications over a Campus-Scale Testbed}

\renewcommand{\shorttitle}{}
\renewcommand{\headeright}{}

\author{ Md Shariful Islam$^1$, Joaquin Chung$^1$, Ely Marcus Eastman$^{1,2}$,\\
\textbf{Robert J. Hayek$^1$, Prem Kumar$^2$, and Rajkumar Kettimuthu$^1$}\thanks{Corresponding authors: Md. Shariful Islam (mislam@anl.gov) and Rajkumar Kettimuthu (kettimut@anl.gov, Project PI).}\\
$^1$Data Science and Learning Division, Argonne National Laboratory, Lemont, IL 60439 USA \\
$^2$Northwestern University, Evanston, IL 60208 USA}

\begin{document}
\maketitle
\begin{abstract}
To fulfill their promise, quantum networks must transform from isolated testbeds into scalable infrastructures for distributed quantum applications. In this paper, we present a prototype orchestrator for the Argonne Quantum Network (ArQNet) testbed that leverages design principles of software-defined networking (SDN) to automate typical quantum communication experiments across buildings in the Argonne campus connected over deployed, telecom fiber. Our implementation validates a scalable architecture supporting service-level abstraction of quantum networking tasks, distributed time synchronization, and entanglement verification across remote nodes. We present a prototype service of continuous, stable entanglement distribution between remote sites that ran for 12 hours, which defines a promising path towards scalable quantum networks.
\end{abstract}

\keywords{quantum networks, entanglement, polarization, timing, synchronization, tomography, fidelity}

\section{Introduction} \label{sec:intro}
Quantum networks promise to enable ultra-secure communication~\cite{ben2014quantum}, distributed quantum computation~\cite{wehner2018quantum}, and quantum-enhanced metrology~\cite{komar2014quantum,kimble2008quantum}. A growing foundation of physics experiments demonstrate this potential, including high-fidelity quantum teleportation over deployed fiber~\cite{Valivarthi2020}, distributed synchronization solutions~\cite{Valivarthi2022,Kapoor2023,Burenkov2023}, and entanglement distribution over urban networks~\cite{kucera2024demonstration}. Unlocking quantum networking's full potential requires seamless integration of multiple quantum networking functionalities into systems capable of providing reliable networking services. However, fundamental physical constraints create unprecedented challenges. Unlike classical systems, quantum information decoheres rapidly under environmental noise, requires unprecedented time synchronization, and relies on probabilistic entanglement generation. These reasons make the coordination of network functions of fragile quantum states with low timing jitter far more complex than conventional networking~\cite{Pompili2022,Illiano2022}.

In recent years, we have witnessed a growing body of work evolving quantum communication experiments into quantum networking systems~\cite{alshowkan2021reconfigurable,alshowkan2025resilient,monga2023quant,yu_two-level_2025,delledonne2025operating,zhao2025quantum,zhang2025switchqnet}. 
This transition requires the integration of multiple quantum networking functionalities into a comprehensive system capable of executing quantum-networked applications.  
Alshowkan et al., demonstrated a reconfigurable quantum network using campus fiber with GPS synchronization~\cite{alshowkan2021reconfigurable}, and later software-defined mesh networking with White Rabbit synchronization achieving sub-nanosecond precision~\cite{alshowkan2025resilient}. 
Yu et al.~\cite{yu_two-level_2025}, proposed a two-level framework for real-time (node level) and non-real-time (network level) task scheduling. This framework was implemented in the QUANT-NET testbed to demonstrate periodic calibration routines of trapped-ion systems and fundamental quantum networking functions (e.g., single photon generation and Bell State measurement).
Delle~Donne et al.~\cite{delledonne2025operating} developed QNodeOS, an operating system that allows end-nodes to schedule application tasks on a quantum network. QNodeOS was demonstrated over a short-range ($<3$~m), point-to-point link using NV-center end-nodes executing delegated quantum computation.

Despite experimental advances and network testbed demonstrations, a significant gap remains in realizing an operational quantum network.
Production-grade quantum networks are expected to operate autonomously, and with high availability over long periods of time.
In this paper, we present the design, implementation, and evaluation of the Argonne Quantum Network (ArQNet) orchestrator, which we use to demonstrate continuous entanglement distribution over a multi-node campus-scale quantum networking testbed.
To achieve this, we create routines for the periodic calibration of an entangled-photon source (EPS) and polarization analyzers (PAs).
We orchestrate the execution of two-photon interference (TPI) and quantum state tomography (QST) experiments. In particular, we automate the TPI experiment to execute over a 12~hour period. The automated TPI experiment continuously monitors relavent performance indicators and triggers link recalibration when any fringe visibility drops below its preset threshold, restoring values to within acceptable operating range.

The rest of this paper is organized as follows. Section~\ref{sec:background} outlines the key motivation behind this work. Section~\ref{sec:design} details the design of the ArQNet orchestrator. Section~\ref{sec:implement} describes the implementation of our orchestrator covering synchronization, entangled-photon source calibration, and polarization drift compensation, as well as the campus-scale testbed used for evaluation. Section~\ref{sec:results} presents the  TPI and QST results for both co-located and fully-remote network configurations, along with a 12-hour continuous, stable entanglement distribution prototype service. Section~\ref{sec:conclusion} concludes with a discussion of scalability and future extensions.
\section{Motivation} \label{sec:background}
The core function of a quantum network is to deliver entanglement between remote users across metropolitan distances. Realizing this requires precise coordination amongst quantum devices, i.e., EPS, Bell state analyzers (BSA), and quantum memories; all operating under tight timing constraints. Similar to classical networks that employ software-defined orchestration for managing disaggregated functions~\cite{nfv-sdn-survey2019}, quantum networks must integrate diverse physical subsystems through a unified control framework.

Accurate synchronization is fundamental for reliable entanglement distribution. Early implementations relied on ad-hoc time alignment and localized clock recovery~\cite{wang2012fast}, which limits scalability. Even campus-scale systems with GPS or White Rabbit timing~\cite{alshowkan2021reconfigurable,alshowkan2025resilient} lack coordinated orchestration, leading to measurement uncertainty and unstable correlations. Therefore, maintaining deterministic clock distribution and automated calibration is essential for reproducible performance across distributed quantum nodes.

Beyond precise timing, network-level orchestration is required to coordinate device calibration, measurement scheduling, and data collection. Without automation, long-duration entanglement distribution services are prone to drift in polarization, phase, and photon count rates. A unified orchestration framework that integrates synchronization, calibration, and measurement control can enable the continuous and stable operation of photonic quantum networks, supporting the scalable entanglement distribution across remote sites.



\section{Architecture Design} \label{sec:design}

\begin{wrapfigure}{r}{0.48\textwidth}
    \centering
    \includegraphics[width=.46\textwidth]{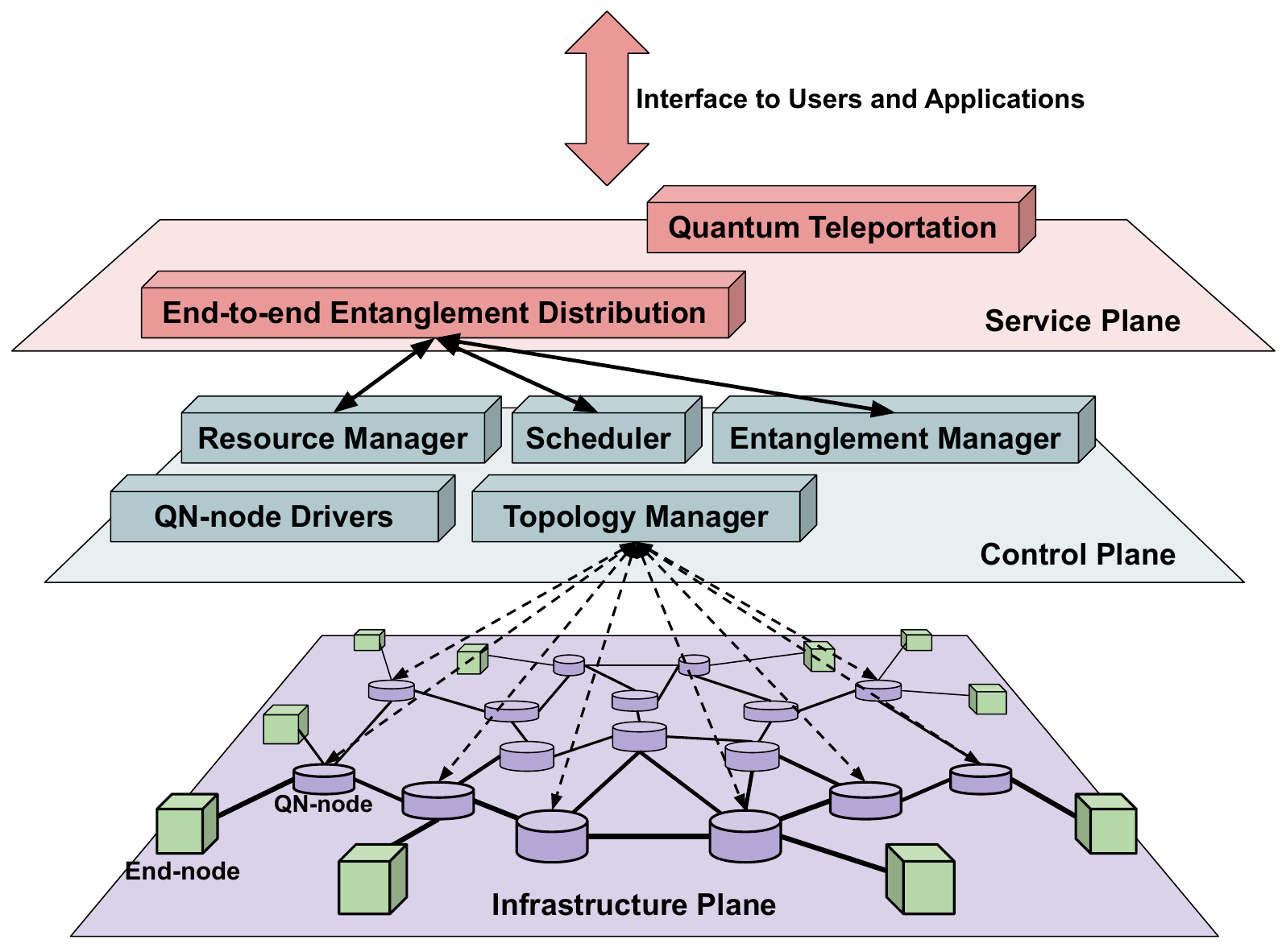}
    \caption{Architecture of a quantum network using a plane abstraction.}
    \vspace{-2ex}
    \label{fig:plane-qnet}
\end{wrapfigure}

Fig.~\ref{fig:plane-qnet} illustrates a quantum network organized using a three-plane abstraction: the infrastructure plane at the bottom represents all physical devices and fiber connections; the control plane in the middle captures fundamental network functions that allow automated operation of quantum network; and the service plane composes user-facing services accessed through standard interfaces.
We adopt this architecture to design the ArQNet orchestrator. 

The infrastructure plane is comprised of the hardware components, i.e., comprises quantum networking devices, all-optical switches, and quantum end-nodes. 
Typically, commercial optical cross-connects are used as all-optical switches, while micro-controllers or low-latency field programmable gate array (FPGA) boards are employed to configure and control quantum devices.
Moreover, quantum devices require very precise synchronization systems as elaborated in Section~\ref{sec:implement}.
The infrastructure plane also comprises the communication media, i.e., optical fibers and free space, connecting quantum devices.
It provides interfaces to allow the control plane to define device configuration parameters and establish on-demand connectivity between devices.
For the ArQNet orchestrator, we are mainly concerned with the quantum network devices involved in entanglement generation processes (purple nodes in Fig.~\ref{fig:plane-qnet}) and we assume that end-nodes have quantum-memories that store their corresponding of the entangled pair.

The control plane is comprised of network functions, a set of operations that collectively achieve a specific objective within the larger network architecture.
Similar to classical networks, a quantum network must perform functions for resource management, scheduling, and topology management in order to provide entanglement distribution services to users. 
However, these functions must be adapted to accommodate the constraints of quantum information.
Moreover, quantum networks require one non-classical network function: entanglement management.
To execute its functions, the control plane consumes the interfaces provided by the infrastructure plane, while it also exposes interfaces to the service plane for end-to-end service composition.
Since near-term quantum networks are likely to be of small scales, we consider a centralized control plane in line with software-defined networking (SDN) principles.

The service plane composes end-to-end services such as end-to-end entanglement distribution and quantum teleportation and it provides these services to users or applications. 
For example, two users that want to securely communicate using quantum key distribution (QKD) can request continuous distribution of entangled pair at a certain rate and for a determined duration.
Similarly, a distributed quantum computing application may request entangled pairs on demand with a specific fidelity threshold to either execute remote two-qubit gates or teleport quantum data.
In our current design, we are mainly concerned with the entanglement distribution service.

\section{Implementation} \label{sec:implement}
\begin{wrapfigure}{r}{0.5\textwidth}
    \centering
    \includegraphics[width=.48\textwidth]{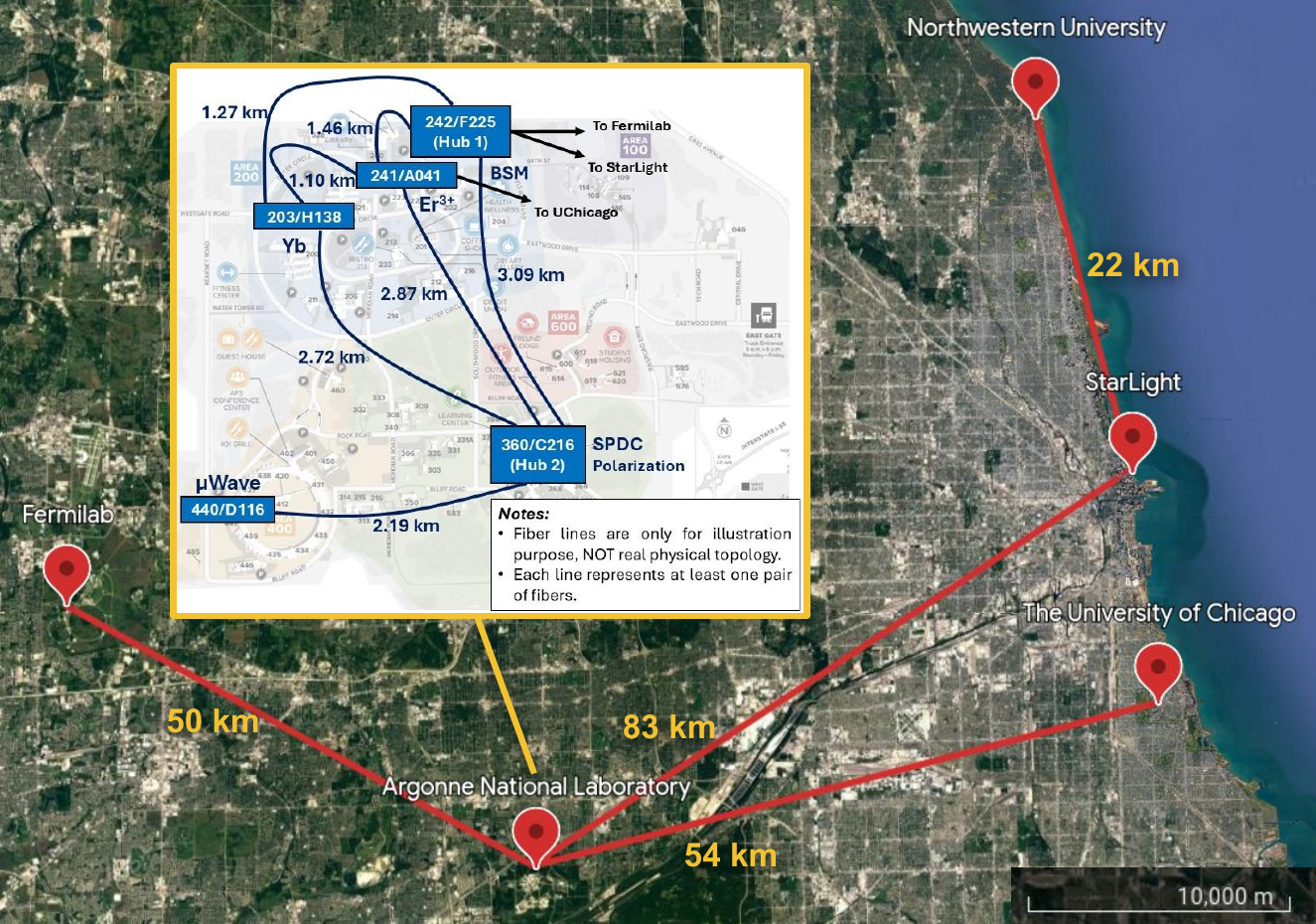}
    \caption{Dark fiber links between Argonne and partner institutions in the great Chicago area. The inset shows Argonne's quantum network testbed (ArQNet).}
    \vspace{-2ex}
    \label{fig:arqnet-topo}
\end{wrapfigure}

We implemented our design on the ArQNet testbed, a campus-to-metro scale heterogeneous quantum networking platform that enables research, development, and validation of the building blocks of scalable quantum networks. 
The network currently has five sites connected in a partial mesh topology with locations at buildings 203, 241, 242 (main hub), 360 (secondary hub), and 440, which are connected via dark fiber links (see inset in Figure~\ref{fig:arqnet-topo}). 
Additionally, ArQNet is connected to partner institutions in the great Chicago area such as Fermilab, the University of Chicago, Northwestern University, and StarLight (Internet Exchange Point) via dark fiber.
The deployment integrates entangled-photon distribution, radio-over-fiber clock delivery, and modular device agents to support synchronized quantum interference and tomography across remote quantum nodes.

\subsection{Experimental Testbed}

\begin{figure*}[htb]
    \centering    \includegraphics[width=0.8\textwidth]{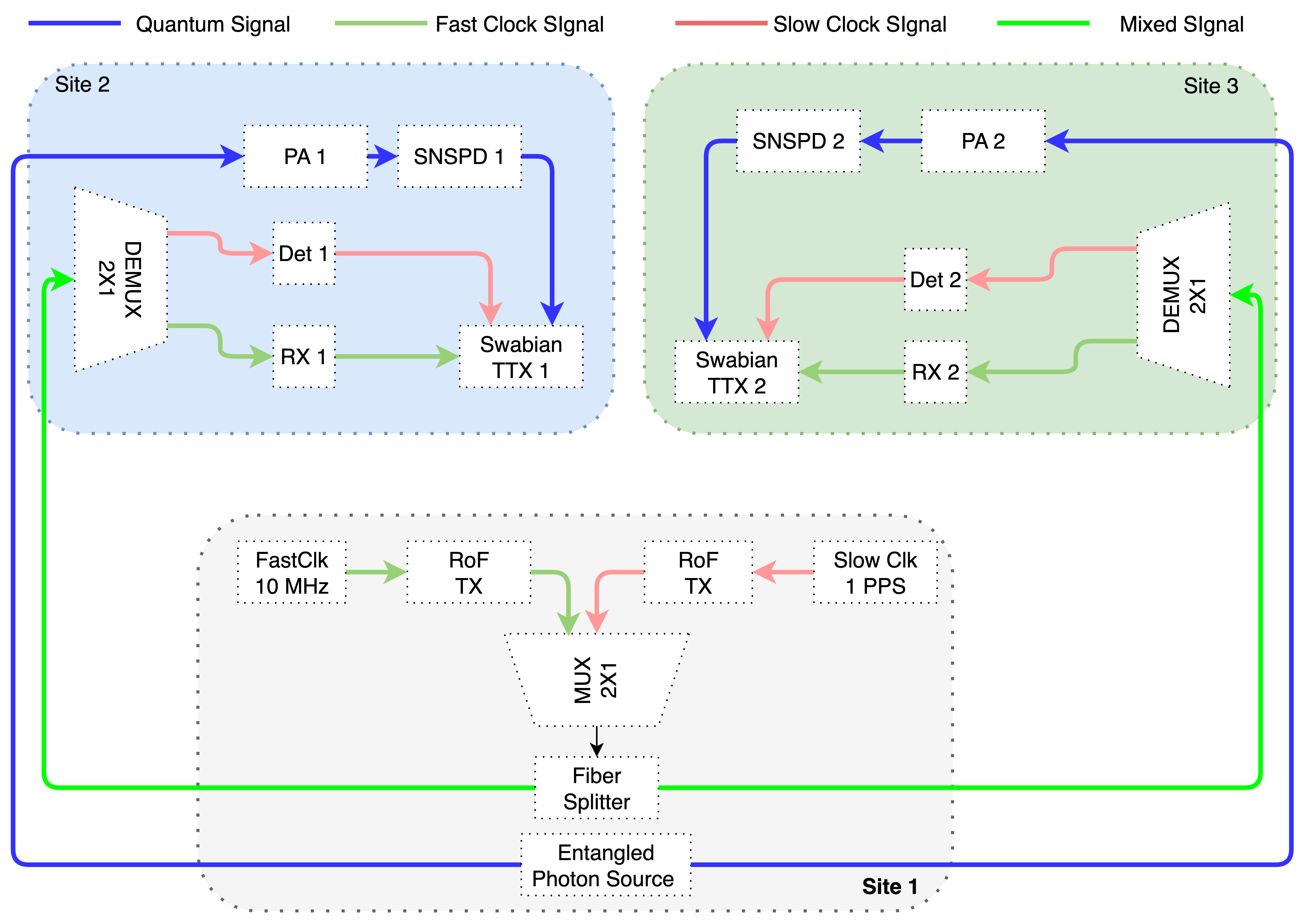}
    \caption{\label{fig:ArQNet_Testbed}
    Schematic diagram of the experimental setup for distributing entangled photons and synchronized clock signals across two remote quantum network nodes (Site~2 and Site~3) from a central entangled photon source (Site~1). A multiplexed (MUX 2$\times$1) fiber output carries both clocks (green and red) via radio-over-fiber (RoF) transmission. The fast (10~MHz) and slow (1~PPS) clock signals are combined at Site~1 and distributed over the same optical fiber to both remote sites, where demultiplexers (DEMUX 2$\times$) separate the 10~MHz and 1~PPS clock. The quantum signal is directed via dedicated fiber channels to the remote sites. At each site, the quantum signal is directed to a polarization analyzer (PA), followed by detection via SNSPDs. The clock signals are routed through receivers (RX1, RX2) and detection modules, with Swabian Time Tagger Ultra (TTX) units logging all time-stamped events for synchronized measurement and correlation. The setup supports automated quantum networking experiments including remote two-photon interference and quantum state tomography.
    }
    
\end{figure*}

Figure~\ref{fig:ArQNet_Testbed} shows the configuration of the ArQNet testbed for this implementation, depicting the three independent sites that comprise our infrastructure plane. Site~1 serves as the central hub, hosting a commercial EPS (NuCrypt EPS-1000-W) that distributes quantum light to two remote quantum nodes at Sites~2 and 3 through dedicated fiber-optic links. To maintain precise temporal synchronization across the distributed network, a radio-over-fiber (RoF) distribution system co-propagates $10$~MHz and $1$~PPS timing references alongside the quantum channels through the same fiber infrastructure. At each remote site, dense wavelength-division multiplexing filters separate these timing signals and feed them directly to Swabian Time Tagger Ultra units for high-resolution time stamping of clock pulses. The quantum channels terminate at remotely controllable PA's (NuCrypt PA-1000) coupled to superconducting nanowire single-photon detectors (SNSPDs) from Quantum Opus.

The EPS architecture employs a Sagnac interferometer loop configuration built around a periodically poled lithium niobate (PPLN) waveguide. A pulsed laser operating at $1550.1$~nm (corresponding to ITU channel~34 in the telecommunications C-band) with adjustable repetition rates spanning $50$ to $1250$~MHz pumps the nonlinear crystal to initiate spontaneous parametric down-conversion (SPDC) processes within the interferometric loop geometry. This configuration generates polarization-entangled photon pairs whose \emph{signal} and \emph{idler} components are spectrally positioned on ITU-T grid channels symmetrically distributed around the pump wavelength. The EPS-1000-W system provides three selectable wavelength channel pairs: $(\mathrm{CH}46,\mathrm{CH}22)$, $(\mathrm{CH}45,\mathrm{CH}23)$, and $(\mathrm{CH}44,\mathrm{CH}24)$, enabling compatibility with standard wavelength-division multiplexing telecommunications infrastructure.

For comprehensive quantum state characterization, each PA-1000 unit implements programmable polarization measurements through a precisely calibrated arrangement of four variable wave plates positioned with optical axes oriented at $45^\circ$ relative to one another, followed by a fixed linear polarizer. This motorized configuration enables rapid, software-controlled polarization basis rotations necessary for both TPI visibility measurements and the complete set of projective measurements required for full quantum state tomography reconstruction. 

We implemented agents for each physical device, i.e., the EPS, PA, and TTU, in our infrastructure plane. These agents combine the physical interface of the device into a uniform API which can be controlled using gRPC. The control plane can then issue commands to stage calibrations and timing actions, with deterministic sequencing, bounded timeouts/retries, and structure provenance logs. These actions include but are not limited to EPS attenuation sweeps, clock locking, and wave-plate scanning. This implementation decouples the control plane logic from hardware differences, enabling identical re-runs, safe recoveries, and allows the same code to target co-located or remote nodes without modification.

\subsection{Control Plane Implementation}
Our control plane is implemented as a centralized orchestrator capable of issuing commands to the infrastructure plane for executing timing and calibration routines, as well as allowing users to automate remote entanglement distribution experiments.
The following subsections provide details on the timing and calibration routines implemented for this work.

\subsubsection{Remote Site Synchronization}

\begin{wrapfigure}{r}{0.5\textwidth}
    \centering
    \includegraphics[width=0.48\textwidth]{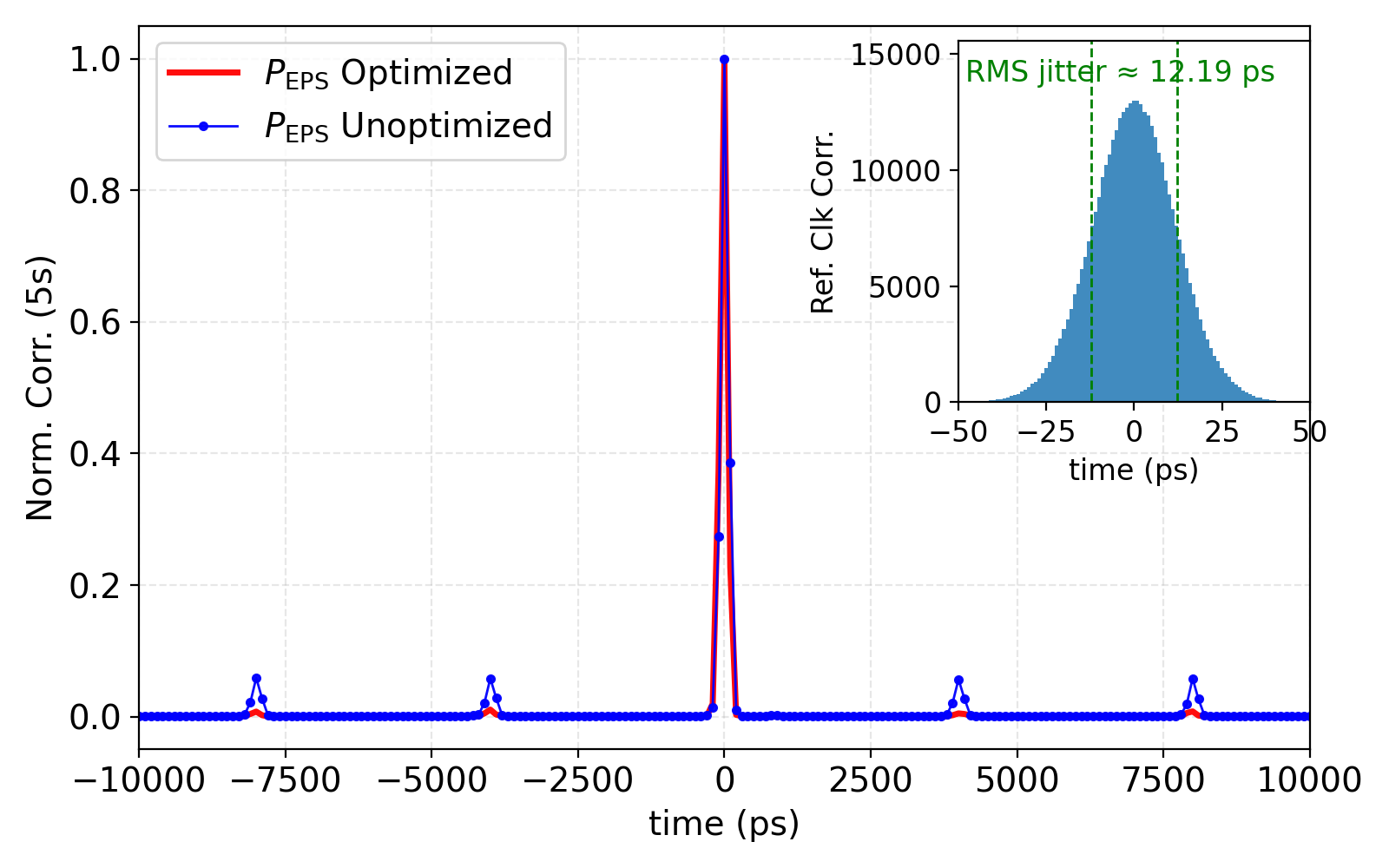}
    \caption{Improvement in coincidence-to-accidental ratio before (blue) and after (red) EPS optimization (main panel). Inset: correlation of $10$~MHz reference clock pulses at the two remote nodes, confirming precise timing alignment for photon detection.}
    \vspace{-4ex}
    \label{fig:correlation}
\end{wrapfigure}
Optical quantum networks require picosecond-level synchronization to maintain entangled correlations across fiber links. Nanosecond-level timing drift can destroy quantum interference, decorrelate detection events, and undermine distributed protocols such as teleportation and QST.

We implemented a hybrid synchronization architecture for two TTU devices deployed across separate sites. 
In the context of a fully-mature quantum network architecture, our sites can be seen as quantum nodes or Q-Nodes as defined in~\cite{ieqnet-arch2022,monga2023quant}.
Our system integrates: (i) a 10~MHz clock for frequency locking, (ii) a 1~PPS signal for absolute alignment, and (iii) Network Time Protocol (NTP) for protocol coordination. Timing signals are co-distributed over dedicated fiber, with the 10~MHz signal locking the TTU oscillator and the PPS signal establishing a common reference. TTU agents use the \texttt{setReferenceClock} function from Swabian to lock to these references, with recovery logic automatically retrying failed synchronization attempts.

\begin{wrapfigure}{r}{0.5\textwidth}
    \centering
    \includegraphics[width=0.48\textwidth]{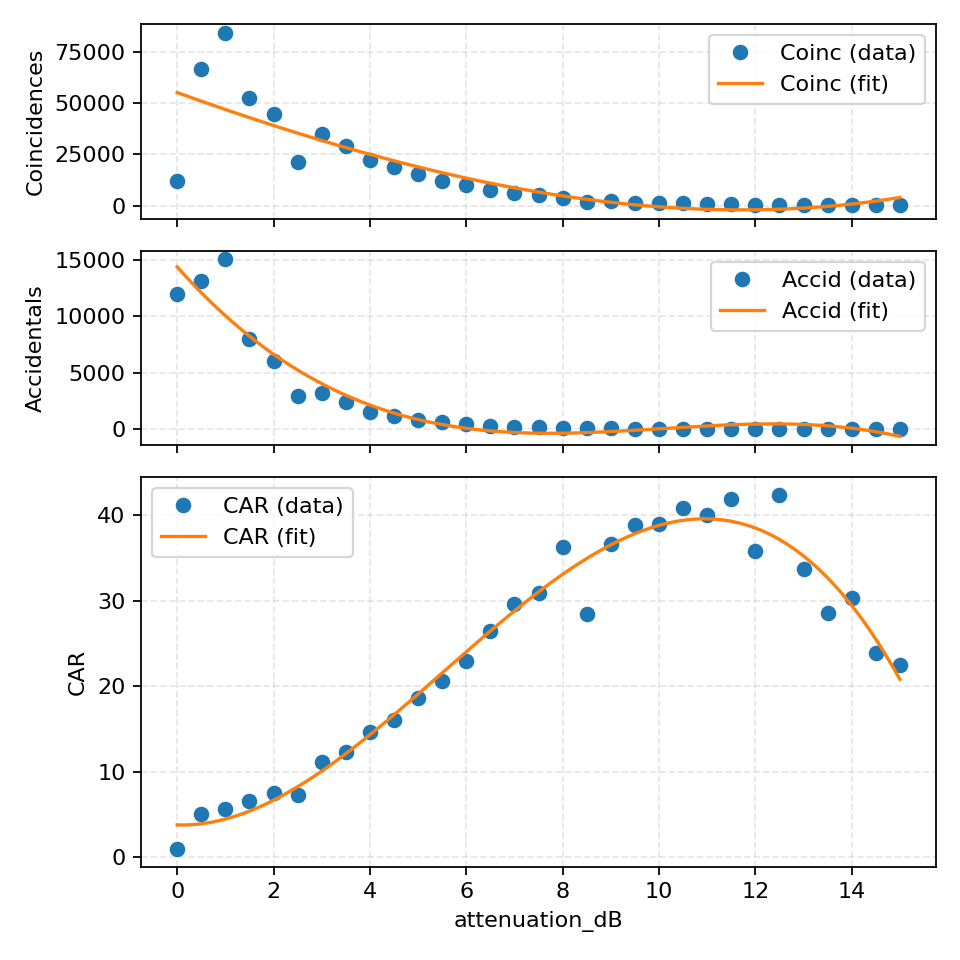}
    \caption{EPS calibration versus pump attenuation. Coincidence (C), accidental (A), and coincidence-to-accidental ratio (CAR). The operating point (red marker) corresponds to $0.85\,\mathrm{CAR}_{\max}$, chosen for stability.}
    \vspace{-10ex}
    \label{fig:EPS_Calib}
\end{wrapfigure}

The central orchestrator establishes gRPC channels to each TTU agent, initializes Swabian's \texttt{SynchronizedMeasurements} context, and launches coordinated acquisition. While NTP provides only millisecond-level coordination, post-processing alignment corrects residual offsets using recorded 1~PPS timestamps. Time streams are collected for 5~s, PPS events extracted from both TTUs, and nearest-neighbor alignment performed via KD-tree search. The median offset is applied during stream merging with \texttt{mergeStreamFiles()}, producing a unified, temporally aligned dataset for coincidence detection.

Figure~\ref{fig:correlation} shows signal–idler correlations before (blue) and after (red) EPS calibration. Without calibration, multipair noise at high pump power and dark counts at very low pump increased accidental coincidences, reducing CAR. Optimization suppresses these effects and enhances fringe visibility. The inset displays cross-correlations of the distributed $10$~MHz reference clocks at the remote nodes, used for time-stamping single-photon events. The measured RMS jitter of $12.19$~ps confirms highly stable clock distribution, a prerequisite for reliable coincidence measurements.

\subsubsection{EPS Calibration}
Figure~\ref{fig:EPS_Calib} shows the calibration of the entangled photon source (EPS) as a function of pump attenuation. Coincidence ($C$) and accidental ($A$) counts increase initially, and then decrease with further attenuation. The coincidence-to-accidental ratio (CAR) exhibits a broad parabolic dependence, reaching a maximum of 47 at 12.5 dB. For stable operation, we selected an operating point at $0.85\times\,\mathrm{CAR}_{\max}$ near 7.5 dB, corresponding to $C=4980$ cps and $A =161$ cps. This lies on a stable region of the CAR curve, providing robustness against pump fluctuations during extended runs.

\subsubsection{Automated Polarization Drift Compensation}

Polarization drift in deployed fiber networks poses a significant challenge to the stability of quantum interference and state tomography experiments involving polarization-encoded optical qubits. Environmental changes such as temperature variation and mechanical stress on fibers can rotate the polarization state of photons unpredictably, degrading interference visibility and reducing the fidelity of reconstructed quantum states. To address this, our orchestrator includes an automated polarization compensation routine that enables remote and autonomous realignment of polarization settings.

The PA implements a single-qubit \emph{projective polarization measurement} by realizing an arbitrary rotation with a train of four variable waveplates (WP0–WP3) followed by a fixed linear polarizer. For a desired analysis state $\ket{b} \in \{\mathrm{H},\mathrm{V},\mathrm{D},\mathrm{A},\mathrm{R},\mathrm{L}\}$, the local controller sets the waveplate retardances so that the net unitary $U(\mathrm{WP0\text{--}3})$ rotates $\ket{b}$ to $\ket{H}$; the polarizer then transmits $\ket{H}$, yielding a count rate proportional to the projection probability
\[
p_b \;=\; \lvert\langle b \vert \psi\rangle\rvert^2 \;=\; \lvert\langle \mathrm{H} \vert U \vert \psi\rangle\rvert^2,
\]
the quantum analogue of Malus’s law \cite{hecht_optics}. The calibration procedure in Algorithm~\ref{alg:pol_drift}, adapted from~\cite{wang2012fast}, establishes a consistent mapping between mechanical waveplate angles and polarization states on the Poincar\'e sphere. To align a PA to a desired measurement basis, an optical alignment signal prepared in that basis is transmitted from the source to the receiving node. The waveplates are sequentially adjusted to minimize the detected photon counts, thereby identifying the orthogonal polarization state corresponding to the transmitted field. The target basis is then obtained through controlled rotations of the waveplates. For the diagonal basis $\ket{D}$, WP3 is fixed at quarter-wave retardance to correct the phase between $\ket{H}$ and $\ket{V}$, while WP2 is tuned to locate the minimum count. Final alignment across both $\ket{H}$ and $\ket{D}$ bases is achieved by iteratively adjusting WP0–WP2 until a configuration minimizing counts for both states is found. 


\begin{figure}[t]
    \centering
\includegraphics[width=\linewidth]{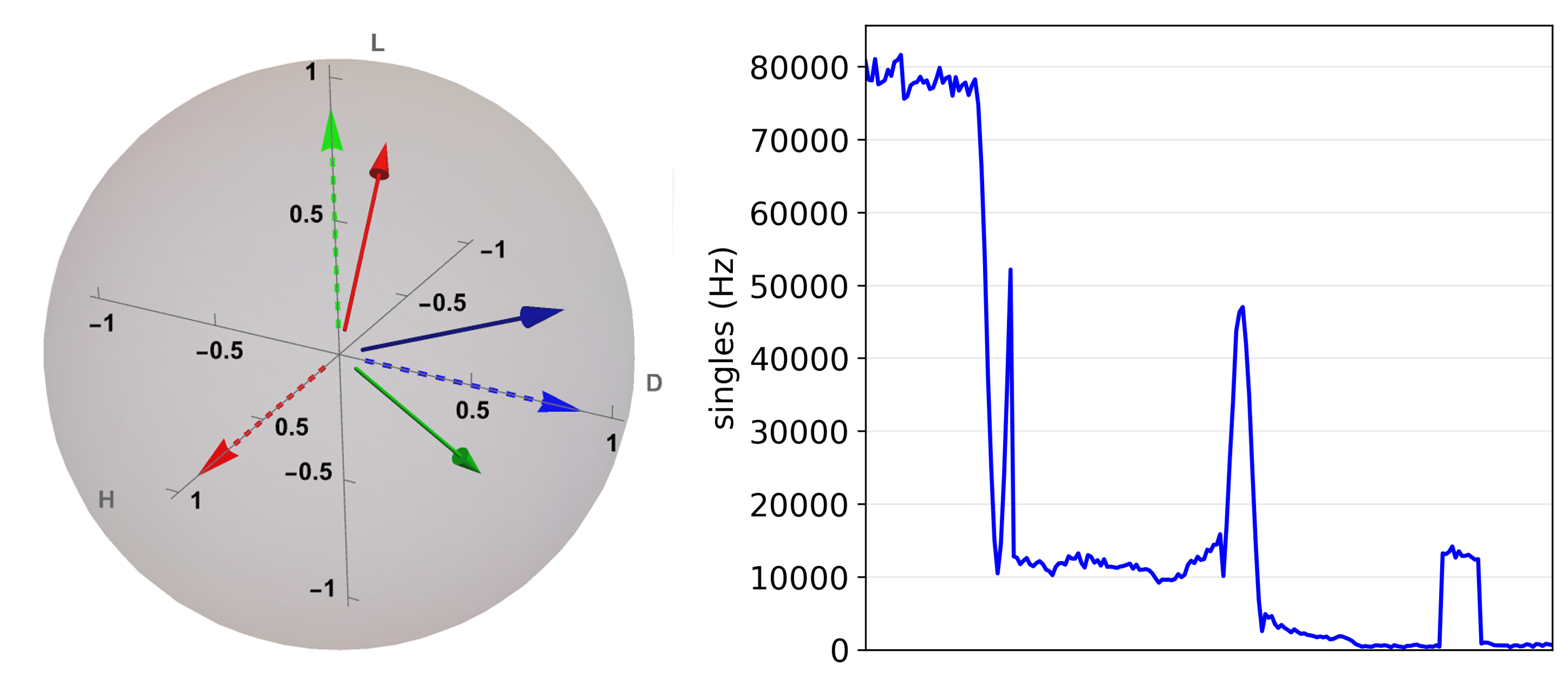}
\caption{Left: Poincare Sphere representation of the misaligned measurement bases at the end of the fiber link (solid) and the ideal measurement bases (dashed) defined by the source's polarizing beam splitter. Right: Singles counts measured by the time tagger during the alignment procedure. A minimum is found near the dark count level of the detectors.}

    \label{fig:Pol_Drift_Adjusted2}
\end{figure}


The polarization drift compensation routine (see Algorithm~\ref{alg:pol_drift} in the Appendix) is initiated by the orchestrator at the beginning of each experimental run to maintain alignment. While ArQNet's underground fiber links remains stable enough to reuse previously calibrated waveplate baselines, the system retains the ability to re-align dynamically when needed. Each polarization analyzer functions as an independent agent, enabling parallel, symmetric correction across nodes with low-latency, platform-independent communication. This software-defined approach converts a physical-layer challenge into a control-layer solution, replacing manual realignment with automated active feedback. As a result, the orchestrator supports robust, repeatable quantum experiments over distributed infrastructure, even under varying environmental conditions.

\subsection{Service Plane Implementation}
We implement a prototype service of continuous, stable entanglement distribution between two remote sites, by orchestrating TPI measurements over a period of 12 hours (see Algorithm~\ref{alg:long_run} in the Appendix). 
After an initial calibration of the EPS and PAs as described in the sections above, the service collects two-photon interference fringes in the $\ket{H}$, $\ket{V}$, $\ket{L}$, and $\ket{R}$ bases, and calculate the average visibility across all four basis measurements $V_{th}$ to use as a threshold for network recalibration. 
We measure TPI every hour, calculate the average visibility $V$, and compare it to the initial value $V_{th}$. 
When $V < V_{th}$, the orchestrator triggers a realignment procedure.
\section{Results and Discussion} \label{sec:results}

\begin{figure*}[tb]
    \centering
\includegraphics[width=\linewidth]{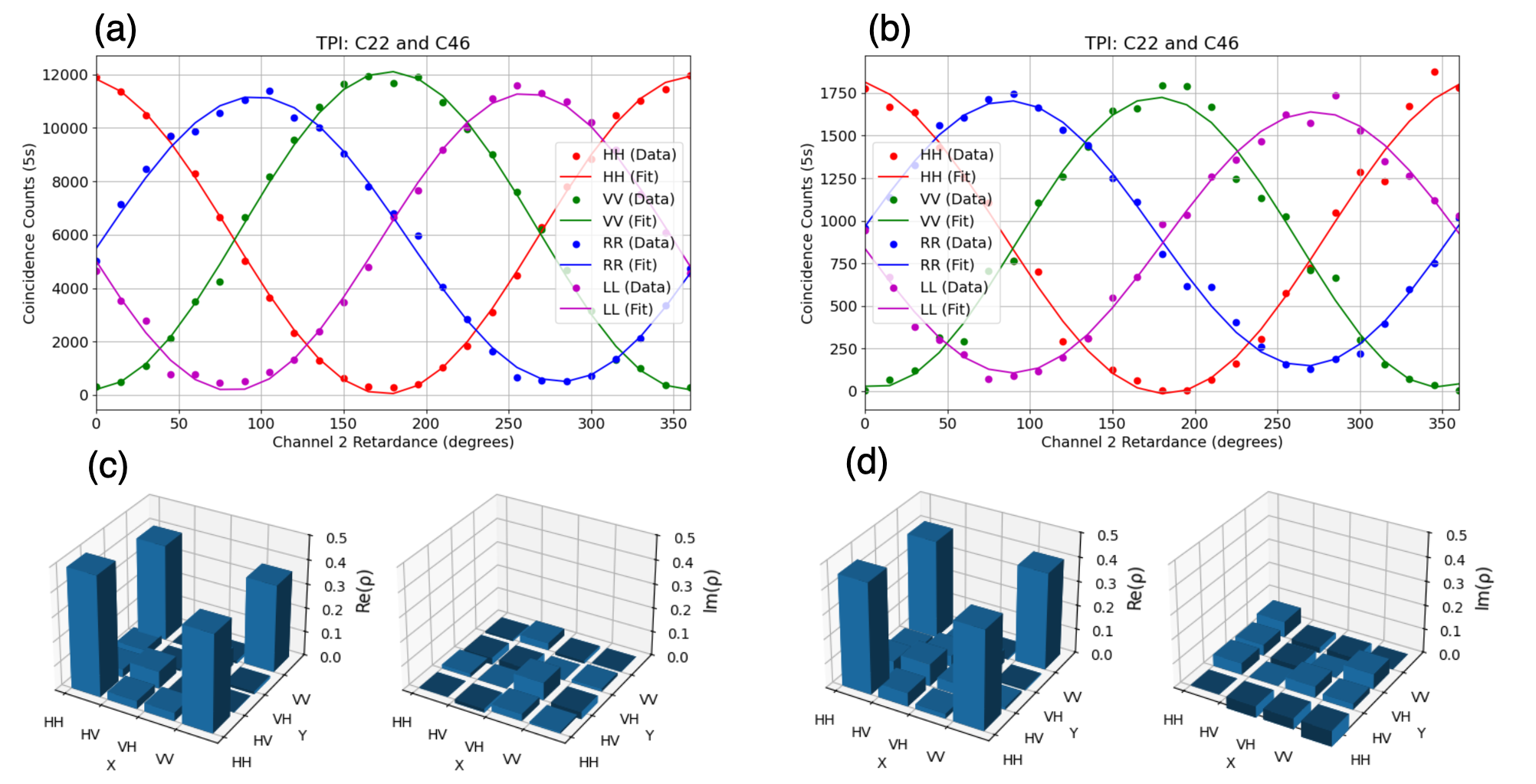}
\caption{Two-photon interference and quantum state tomography under different network configurations. (a) Co-located EPS and Detector~1 with Detector~2 remote. (b) EPS at Site~1 with both detectors remote (Sites~2 and 3). (c) QST corresponding to (a), showing fidelity $F \approx$ 82.56. (d) QST corresponding to (b), showing fidelity $F \approx$ 83.30. Highest interference visibilities were computed in HH bases for both cases and are $V \approx$ 95\% and 92\% for (a) and (b), respectively, confirming robust entanglement distribution across distributed quantum nodes.}
    \label{fig:TPI-QST}
\end{figure*}


To demonstrate the functionality of our orchestrator and control plane we observe two-photon interference between measurement nodes and determine the density matrix of our distributed entangled state.
During two-photon interference experiments, the orchestrator directs polarization analyzers to cycle through predefined polarization settings while measuring the variations in coincident detection events between remote nodes.
To fully verify the entangled nature of our quantum state, fringes are observed in multiple different polarization bases.
Our orchestrator automatically collects the synchronized coincident counts, fits and plots interference fringes, and reports visibility statistics without any input from the user.
We can characterize the entangled state's density matrix by collecting coincidence counts across 36 different pairs of projective polarization measurements, requiring at least 144 hardware reconfigurations.
All data are timestamped and processed to reconstruct the density matrix using a least-squares estimation.


Figure~\ref{fig:TPI-QST} summarizes the two-photon interference and quantum state tomography (QST) results under two network configurations.
In (a), Detector~1 is co-located with the EPS while Detector~2 is remote over $\sim$2.19~km of fiber.
Clear interference fringes are observed across all bases with visibility $V_{HH} \approx$ 95\%.
Full visibility statistics are found in Table \ref{tab:vis_stats}.
The decrease in visibility for the circularly polarized bases can be attributed to incomplete alignment of the phase between the horizontal and vertical polarization states.
The corresponding QST in Figure \ref{fig:TPI-QST}(c) yields fidelity $F \approx$ 82.6\% with the target Bell state $\ket{\Phi^+} = \left( \ket{HH} + \ket{VV}\right)/\sqrt{2}$.
In Figure \ref{fig:TPI-QST}(b), both detectors are located remotely from the EPS at Sites~2 and 3, separated respectively by $\sim$3.19~km and $\sim$5.28~km of deployed fiber.
In this arrangement, the polarization measurement bases at each detector site must be aligned to those defined by the polarization beam splitter in the EPS at the source node.
Furthermore, the orchestrator must maintain synchronization between two asymmetric links of fiber experiencing uncorrelated noise.
Despite these additional challenges, we measure high-quality interference with $V_{HH} \approx$ 92\%, and the QST in Figure \ref{fig:TPI-QST}(d) has a fidelity $F \approx$ 83.3\%, comparable to our result in the co-located case.

\begin{wrapfigure}{r}{0.5\textwidth}
    \centering
    \includegraphics[width=0.48\textwidth]{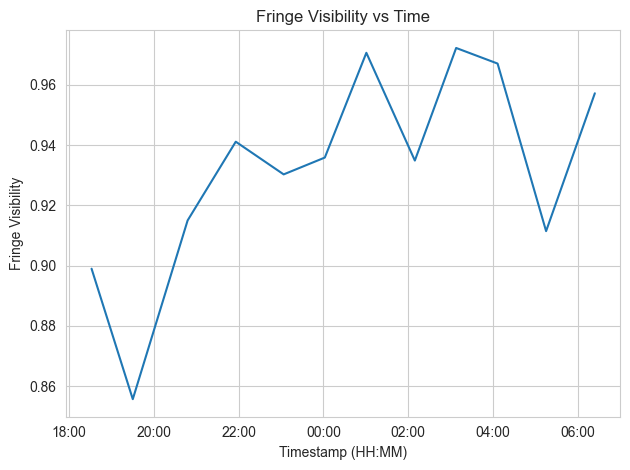}
    \caption{Orchestrated TPI measurement over 12 hours computing average fringe visibility of the $\ket{H}$, $\ket{V}$, $\ket{L}$, $\ket{R}$ polarization bases and reporting average visibility $V$.}
    \label{fig:12hr-experiment}
\end{wrapfigure}

To further quantify the entanglement of our distributed source, we measure the concurrence, which ranges from 1 (fully entangled state) to 0 (fully mixed state).
In the co-located (remote) case we find a concurrence of $C = 0.653$ ($C=0.704$).
These results demonstrate the precision of the ArQNet orchestrator's control plane, and its ability to automatically generate and characterize high-quality entanglement across a campus-scale quantum network testbed with no involvement from the user.
This contrasts with a manually run experiment that requires multiple operators at different 
sites coordinating measurement settings and data collection. Our orchestrator reduces both manpower required to operate the experiment to a single user and experiment run time from over one day to approximately 1.5 hours.
Further improvements in run time can be made with software optimization in the orchestrator.


Figure~\ref{fig:12hr-experiment} presents the evaluation results of our continuous, stable entanglement distribution service prototype.
The experiment started at 18:00 and ran for 12 hours, maintaining above threshold.
We note that the polarization drift compensation only ran once at the beginning of the experiment, because our network is already very stable (all fibers are underground).
Nevertheless, these results provides us with confidence that our orchestrator is capable of turning quantum communication experiments into quantum networking services.

\begin{table}[bhtp]
\caption{Visibilities for Two-Photon Interference Measurements}
\label{tab:vis_stats}
\centering
\begin{tabular}{c c c}
    \toprule
    \textbf{Basis} & \textbf{Colocated} & \textbf{Remote}\\
    \midrule
    $\ket{HH}$ & 95.5\% & 99.8\%\\
    $\ket{VV}$ & 95.4\% & 99.7\%\\
    $\ket{RR}$ & 91.5\% & 86.1\%\\
    $\ket{LL}$ & 92.3\% & 92.3\%\\
    \bottomrule
\end{tabular}
\end{table}
\section{Conclusion} \label{sec:conclusion}

We presented the design, implementation, and evaluation of the ArQNet orchestrator, a quantum network control plane software that integrates distributed timing, polarization stabilization, and modular hardware control into a unified experimental platform. 
By adopting SDN design principles and the three-plane abstraction, the orchestrator (control planes) enables remote coordination of entangled photon sources, polarization analyzers, and time taggers (devices of the infrastructure plane) for the execution of quantum communication experiments (service plane) with minimal manual intervention.  

Our experimental results validate this architecture across colocated–remote and fully remote configurations. EPS calibration improved coincidence-to-accidental ratios to stable operating points, clock distribution over radio-over-fiber achieved sub-20~ps jitter, and automated polarization compensation maintained interference visibility and tomography fidelity over $5$~km of deployed fiber. 
Moreover, we demonstrate a prototype service of continuous, stable entanglement distribution between remote sites capable of maintaining fringe visibility above a preset threshold for 12 hours autonomously.
Together, these results confirm that distributed entanglement generation, two-photon interference, and full quantum state tomography can be reliably executed over campus-scale testbeds.  

This work demonstrates a pathway toward programmable, service-oriented photonic quantum networks. By combining lightweight synchronization with modular orchestration, ArQNet bridges the gap between table-top quantum optics and scalable quantum internet infrastructures. Future extensions will focus on integrating heterogeneous quantum devices, multi-user quantum networking, and fault-tolerant entanglement distribution protocols, further advancing the vision of distributed quantum computing and secure quantum communications.

\medskip
\textbf{Acknowledgments} \par
This material is based upon work supported by the U.S. Department of Energy, Office Science, Advanced Scientific Computing Research (ASCR) program under contract number DE-AC02-06CH11357 as part of the InterQnet quantum networking project.

\textbf{Government License} \par
The submitted manuscript has been created by UChicago Argonne, LLC, Operator of Argonne National Laboratory (``Argonne''). Argonne, a U.S. Department of Energy Office of Science laboratory, is operated under Contract No. DE-AC02-06CH11357. The U.S. Government retains for itself, and others acting on its behalf, a paid-up nonexclusive, irrevocable worldwide license in said article to reproduce, prepare derivative works, distribute copies to the public, and perform publicly and display publicly, by or on behalf of the Government.  The Department of Energy will provide public access to these results of federally sponsored research in accordance with the DOE Public Access Plan. \href{http://energy.gov/downloads/doe-public-access-plan}{http://energy.gov/downloads/doe-public-access-plan.}

\bibliographystyle{IEEEtran}
\bibliography{References}


\appendix
\section{Appendix}
This appendix presents pseudo code for our polarization drift compensation routine and the continuous, stable entanglement distribution service prototype.

\subsection{Polarization Drift Compensation}
\begin{algorithm}
    \caption{Polarization Drift Compensation}
    \label{alg:pol_drift}
    \begin{algorithmic}[1]
        \State $\theta_2$ $\gets \arg\min_\theta(\text{WP2}, s, w, \ket{D})$
        \State $\theta_1$ $\gets \arg\min_\theta(\text{WP1}, s, w, \ket{H})$
        \State $\theta_0$ $\gets \arg\min_\theta(\text{WP0}, s, w, \ket{H})$
        \Function{MinimizeWaveplate}{WP, $s$, $w$, $\ket{b}$}
            \State $\theta_c \gets$ current WP angle
            \For{$\theta \in \{\theta_c - w, \theta_c - w + s, \ldots, \theta_c + w\}$}
                \State Set WP to angle $\theta$
                \State $S(\theta) \gets$ measure singles count rate in $\ket{b}$
            \EndFor
            \State $\theta^* \gets \arg\min_{\theta} S(\theta)$
            \State \Return $\theta^*$
        \EndFunction
    \end{algorithmic}
\end{algorithm}
\newpage
\subsection{Continuous, Stable Entanglement Distribution Service}
\begin{algorithm}[htbp]
\label{alg:long_run}
\caption{Entanglement Distribution Service}
\textit{Note: Polarization drift compensation is performed using Algorithm \ref{alg:pol_drift}.}
\begin{algorithmic}[1]
\State \textsc{Initialize}(EPS, PA1, PA2)
\State
\State $(D_{signal}, D_{idle}) \gets$ \textsc{GetDarkCounts}()
\If{$D_{signal} \notin [D_{min}, D_{max}]$ \textbf{or} $D_{idle} \notin [D_{min}, D_{max}]$}
    \State \textbf{abort} experiment
\EndIf
\State
\State \textsc{CompensatePolarizationDrift}(PA1, PA2)
\State $(max\_car, attenuation) \gets$ \textsc{CalibrateEPS}()
\State
\State $start\_time \gets$ \textsc{Now}()
\While{$(\textsc{Now}() - start\_time) < run\_time$}
    \State $\{\theta, H, V, R, L\} \gets$ \textsc{RunTPI}(PA1, PA2)
    \State $\mathcal{V} \gets$ \textsc{ComputeFringeVisibility}($\theta, H, V, R, L$)
    \State
    \If{$\exists b \in \{H, V, R, L\} : \mathcal{V}_b < threshold$ \textbf{or} $\mathcal{V}_b = \text{NaN}$}
        \State \textsc{CompensatePolarizationDrift}(PA1, PA2)
    \Else
        \State \textsc{Wait}($\Delta t$)
    \EndIf
\EndWhile
\State
\Function{CalibrateEPS}{}
    \State $CAR_{max} \gets 0$
    \For{$\alpha \in \{0, 0.5, 1.0, \ldots, 15.5\}$ dB}
        \State Set EPS attenuation to $\alpha$
        \State $(C, A) \gets$ measure coincidence and accidental
        \State $CAR(\alpha) \gets \frac{C}{A}$
        \State $CAR_{max} \gets \max(CAR_{max}, CAR(\alpha))$
    \EndFor
    \State $CAR_{target} \gets 0.8 \cdot CAR_{max}$
    \State $\alpha^* \gets \arg\min_{\alpha} |CAR(\alpha) - CAR_{target}|$
    \State Set EPS attenuation to $\alpha^*$
    \State \Return $(CAR_{max}, \alpha^*)$
\EndFunction
\State
\Function{ComputeFringeVisibility}{$\theta, H, V, R, L$}
    \For{each basis $b \in \{H, V, R, L\}$}
        \State Fit $I(\theta) = A \sin(B\theta + C) + D$ to counts $b(\theta)$
        \If{$D > 0$}
            \State $\mathcal{V}_b \gets \frac{|A|}{D}$
        \Else
            \State $\mathcal{V}_b \gets \text{NaN}$
        \EndIf
    \EndFor
    \State \Return $\mathcal{V} = \{\mathcal{V}_b : b \in \{H, V, R, L\}\}$
\EndFunction
\end{algorithmic}
\end{algorithm}

\end{document}